\documentclass[aps,prl,superscriptaddress,twocolumn]{revtex4}
\usepackage{amsmath,epsfig,color,amssymb}
\usepackage{graphicx}
\usepackage{graphicx}
\usepackage{dcolumn}% Align table columns on decimal point
\usepackage{bm}% bold math
\usepackage{graphicx}
\usepackage{amsmath, amsfonts, amssymb,mathrsfs}
\usepackage{pstricks}
\usepackage{amsxtra}
\usepackage{amsthm}
\usepackage{natbib}

\definecolor{new}{rgb}{.08,.05,.8}

\def\be{\begin{equation}}
\def\ee{\end{equation}}
\def\bea{\begin{eqnarray}}
\def\eea{\end{eqnarray}}
\newcommand{\ket}[1]{\mbox{$|#1\rangle$}}

\def\be{\begin{equation}}      % with numbering
\def\ee{\end{equation}}
\def\beu{\begin{equation*}}   % without numbering
\def\eeu{\end{equation*}}

\providecommand{\ket}[1]{\left|#1\right\rangle}

\providecommand{\del}{\partial}
\providecommand{\br}{{\bm{r}}}

\providecommand{\bq}{{\bm{q}}}

\begin{document}
\title{Efimov States of Strongly Interacting Photons}
\date{\today}
\author{M. J.  Gullans}
 \altaffiliation[Present address: ]{Department of Physics, Princeton University, Princeton, NJ 08544, USA}
 \altaffiliation{mgullans@princeton.edu}
\affiliation{Joint Quantum Institute, NIST and University of Maryland, College Park, Maryland 20742, USA}
\affiliation{Joint Center for Quantum Information and Computer Science, NIST and University of Maryland, College Park, Maryland 20742, USA}
\author{S. Diehl}
\affiliation{Institut f{\" u}r Theoretische Physik, Universit{\" a}t zu K{\" o}ln, D-50937 Cologne, Germany}
\author{S. T. Rittenhouse}
\affiliation{Department of Physics, The United States Naval Academy, Annapolis, Maryland 21402, USA}
\author{B. P. Ruzic}
\affiliation{Joint Quantum Institute, NIST and University of Maryland, College Park, Maryland 20742, USA}
\author{J. P. D'Incao}
\affiliation{JILA, University of Colorado and NIST, Boulder, Colorado 80309, USA}
\affiliation{Department of Physics, University of Colorado, Boulder, Colorado 80309, USA}
\author{P. Julienne}
\affiliation{Joint Quantum Institute, NIST and University of Maryland, College Park, Maryland 20742, USA}
\author{A. V. Gorshkov}
 \altaffiliation[]{gorshkov@umd.edu}
 \affiliation{Joint Quantum Institute, NIST and University of Maryland, College Park, Maryland 20742, USA}
\affiliation{Joint Center for Quantum Information and Computer Science, NIST and University of Maryland, College Park, Maryland 20742, USA}
\author{J. M. Taylor}
 \altaffiliation{jmtaylor@jqi.umd.edu}
\affiliation{Joint Quantum Institute, NIST and University of Maryland, College Park, Maryland 20742, USA}
\affiliation{Joint Center for Quantum Information and Computer Science, NIST and University of Maryland, College Park, Maryland 20742, USA}
\affiliation{Research Center for Advanced Science and Technology (RCAST), The University of Tokyo, Meguro-ku, Tokyo, 153-8904, Japan}
\begin{abstract}
We demonstrate the emergence of universal Efimov physics for interacting photons in cold gases of Rydberg atoms. 
We consider the behavior of three photons injected into the gas in their propagating frame, where a paraxial approximation allows us to consider them as massive particles.  In contrast to atoms and nuclei, the photons have a large anisotropy between their longitudinal mass, arising from dispersion, and their transverse mass, arising from diffraction.  Nevertheless, we show that in suitably rescaled coordinates the effective interactions become dominated by  $s$-wave scattering near threshold and, as a result, give rise to an Efimov effect near unitarity.  
We show that the three-body loss of these Efimov trimers can be strongly suppressed and determine conditions under which these states are observable in current experiments.  These effects can be naturally extended to probe few-body universality beyond three bodies, as well as the role of Efimov physics in the non-equilbrium, many-body regime.
\end{abstract}
\maketitle

 The problem of classifying the universal properties of few-body systems near unitarity, i.e., a divergence in the two-body scattering length $a$, was first undertaken for the three-body problem by Vitaly Efimov in 1970, who discovered an infinite series of three-body bound states obeying a geometrical scaling relation \cite{Efimov71}.   This discovery served as an important guide to theoretical work in few-body physics in subsequent years \cite{Braaten06}, but the observation of such Efimov trimers in nature remained elusive until pioneering experiments on  cold  atomic gases reported direct signatures of these states in atomic loss spectroscopy \cite{Kraemer06}.  That success  reinvigorated work on the classification problem alluded to above, including in systems other than cold atoms \cite{Nishida13,Kunitski15}.    As a result, recent years have seen the elucidation of many universal properties of $N$-body systems for $N\ge 3$ \cite{Platter04,Hammer07,Ferlaino09,vonStecher09,Mehta09,Schmidt10,Zenesini13,Greene17}, including the many-body, short-time dynamics of Efimov trimers in a unitary Bose gas \cite{Klauss17}.   Despite this progress, Efimov states, as well as larger bound state clusters, are typically associated with large inelastic losses in cold atom systems due to strong three-body recombination.  
These losses generally preclude the study of many-body physics of Efimov trimers such as the formation of a Bose-Einstein condensate of trimers \cite{Braaten03}, and limit efforts to study universal bound states for large $N$.

Recently, it has become possible to achieve strong interactions between single photons by dressing light with strongly interacting Rydberg atoms to form Rydberg polaritons \cite{Pritchard10,Gorshkov11}.  
The resulting photon-photon interactions have been used to study a diverse array of quantum nonlinear optical effects including: single-photon blockade and transistors \cite{Kuzmich12,Peyronel12,Hofferberth14,Tiarks14,Baur14}, two-photon phase gates \cite{Tiarks16,Thompson17,Pohl17,Busche17}, and the formation of one dimensional few-photon bound states \cite{Firstenberg13,Buchler14,Maghrebi15,Gullans16,Buchler16,Liang17,Gullans17}.  Combining these strong interactions with the high degree of available control over the optical and atomic degrees of freedom
 makes these systems a promising platform for exploring non-equilibrium quantum many-body physics  and realizing quantum simulation \cite{Pohl10,Parigi12,Otterbach13,Gorshkov13,Maghrebi15b,Moos15,Sommer15,Grankin16,Schine16,Jia17,Zeuthen17}.  

In this Letter, we show how such systems of interacting photons can lead to the formation of Efimov states of light.  We extend previous work on bound states of photons in Rydberg polariton systems by accounting for the three-dimensional (3D) nature of the photons.  For attractive interactions, these considerations lead to the possibility of forming 3D bound states, of which the Efimov states are the first class of three-body bound states that emerge as the strength of the interactions is increased from zero.  

Crucial to the realization of Efimov states in this system is their low-energy, long-wavelength nature, which leads to their emergence independent of many of the microscopic details of Rydberg polaritons.  We use this property to show that the three-body losses of these Efimov states can be strongly suppressed, allowing for the formation of long-lived Efimov trimers.  We analytically demonstrate that this class of Efimov states have anisotropic spatial wavefunctions due to the anisotropic effective mass of the polaritons.  To prepare these states we take advantage of the fact they propagate in the medium as the three-body limit of an optical soliton.  This property allows them to be distinguished from non-bound states in the system, which will dephase due to dispersive and diffractive effects \cite{Firstenberg13,Liang17}.  Finally, we consider the conditions under which these states can be directly observed in current experiments.

\begin{figure}[t]
\begin{center}
\includegraphics[width = .45 \textwidth]{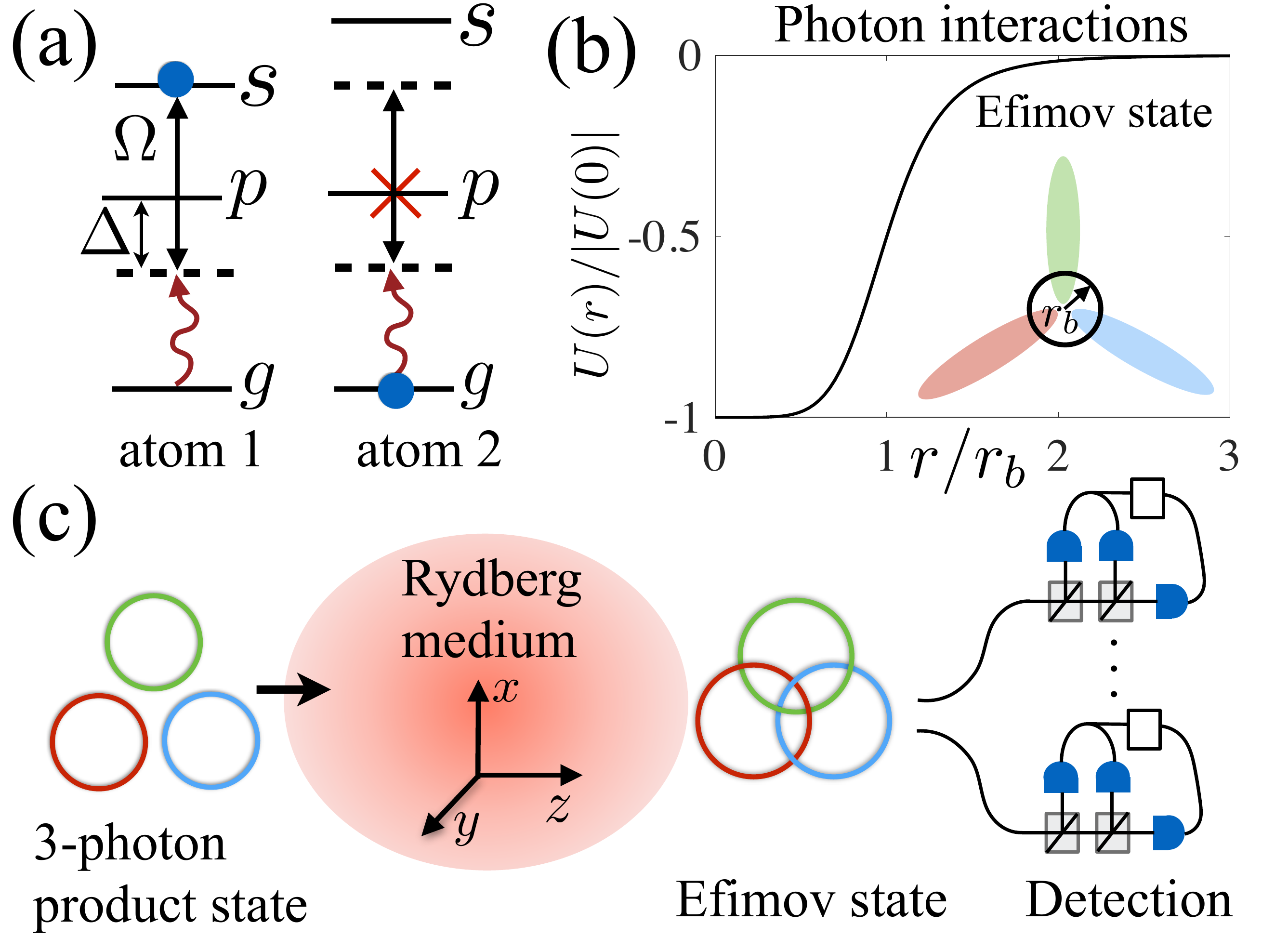}
\caption{(a) Schematic of EIT and the Rydberg blockade effect.  (b) Effective interaction potential for the Rydberg polaritons. (inset) The Efimov states have a large spatial extent compared to the microscopic range of the potential, leading to many simplifications.  (c) The Efimov states emerge in transmission because they are immune to dispersion and diffraction inside the medium.  Spatially resolving multi-mode, three-photon coincidence measurements allow for detailed characterization of these states \cite{Pan12,Liang17}.} 
\label{fig:cartoon}
\end{center}
\end{figure}

The basic configuration to realize interacting photons via  atomic Rydberg states is shown in Fig.~\ref{fig:cartoon}(a).  The Rydberg atoms are dressed with a quantum field of light using electromagnetically induced transparency (EIT).
%The ground state $\ket{g_1}$ is coupled via probe light to an intermediate state $\ket{p}$, which is coupled to the Rydberg state $\ket{s}$ through a classical control field with Rabi frequency $\Omega$. Quantizing the probe field leads to the notion of hybridized excitations of light and matter referred to as dark-state polaritons \cite{}.  
Rydberg-Rydberg interactions lead to the Rydberg blockade effect \cite{Lukin01}, whereby  a single atom (polariton) in the state $\ket{s}$ shifts the  $s$-state of nearby atoms out of resonance, leading to a strong optical nonlinearity.  
To describe the light transmission of this system,  we introduce a bosonic field $\psi(\bm{r})$ associated with the  Rydberg dark-state polaritons.  
 The bare interaction between the Rydberg atoms is given by the van der Waals interaction $V(\br) = -C_6/r^6$, however, the effective interactions between polaritons take the form [see Fig.~\ref{fig:cartoon}(b)]  \cite{Buchler14}
\be \label{eqn:u}
U(\bm{r}) =  \frac{ \alpha V(\bm{r})}{1- \bar{\chi} V(\bm{r})},
\ee
where $\bar{\chi}$ and $\alpha$ are a function of the control parameters and atomic decay rates \cite{supp}.  Here we consider the case $\bar{\chi},C_6>0$ with a large detuning $\Delta$ from the intermediate state so that $U(\bm{r})$ is attractive, non-singular, and conservative. 
$U(\br)$ has a long range van der Waals tail, but saturates to a constant value for $ r \ll r_b$, where the blockade radius is defined as $r_b=|\bar{\chi} C_6|^{1/6}$.  Due to their low-energy, the Efimov states will have a spatial extent much larger than $r_b$ [see inset to Fig.~\ref{fig:cartoon}(b)].  We take advantage of the long-wavelength nature of these states to describe their 
propagation using the effective second-quantized Hamiltonian density
\be \label{eqn:H}
\begin{split}
\mathcal{H}& =  \psi^\dagger(\bm{r})\bigg[ -i \hbar v_g \del_z -  \frac{\hbar^2 \del_z^2}{2 m_z}  - \frac{\hbar^2 \del_\perp^2}{2 m_\perp}  \bigg] \psi(\bm{r}) \\
& + \int d^3r'\, \psi^\dagger(\bm{r}) \psi^\dagger(\bm{r}') U(\bm{r}-\bm{r}') \psi(\bm{r}') \psi(\bm{r}),
\end{split}
\ee
where $v_g$ is the EIT group velocity, $m_{z}$ is the longitudinal  mass arising from dispersive effects, $m_\perp$ is the transverse mass arising from diffraction in the paraxial wave approximation, and $\del_\perp^2 = \del_x^2 + \del_y^2$.  The lowest order correction to this effective Hamiltonian is a short-range, three-body force \cite{Gullans16,Buchler16,Liang17,Gullans17}; however, such  forces typically play a minor role in Efimov physics so we neglect them here  \cite{Wang12,Naidon14b,Naidon14}.

%In related work we have shown that the corrections to this model for low-energy, few-body scattering are strongly suppressed in the regime of large optical depth per blockade radius \cite{Gullans16,Gullans17}, which is the regime considered in this work.

After transforming into a co-moving frame, apart from the anisotropic mass, the effective model has a standard form studied in few-body atomic and nuclear systems.
Furthermore, near threshold, we find that  the anisotropy in the mass can be accounted for by a simple rescaling of the coordinates and the scattering becomes isotropic in the absence of higher partial wave resonances.  This implies that the universal few-body hierarchy, beginning with the Efimov effect will arise near unitarity for such 3D Rydberg polaritons.
%An important distinction from conventional Efimov states is that the wavefunction of the Efimov states in this system becomes anisotropic in the original coordinates.

The  preparation and detection scheme for the Efimov states is illustrated in Fig.~\ref{fig:cartoon}(c). The entrance into the medium acts as a quantum quench  \cite{Firstenberg13,Gullans16}, generating a finite overlap with the Efimov states.  Once they are formed inside the medium, the bound states propagate without distortion, while the scattering states dephase with each other.  As a result, for a sufficiently long medium in the absence of losses, the output will be dominated by the Efimov states.   This effect  has been used previously in 1D Rydberg polariton experiments to directly observe the formation of two and three-body bound states \cite{Firstenberg13,Liang17}.    When there is more than one   bound state in the medium, these states are distinguishable by their spatial structure or  propagation phase through the medium.   To directly probe these states one can use time-resolved, three-photon coincidence measurements to access their longitudinal spatial structure, while  multi-mode spatial resolution can probe their transverse structure \cite{Pan12,Liang17}.  

\begin{figure}[t]
\begin{center}
\includegraphics[width=.48\textwidth]{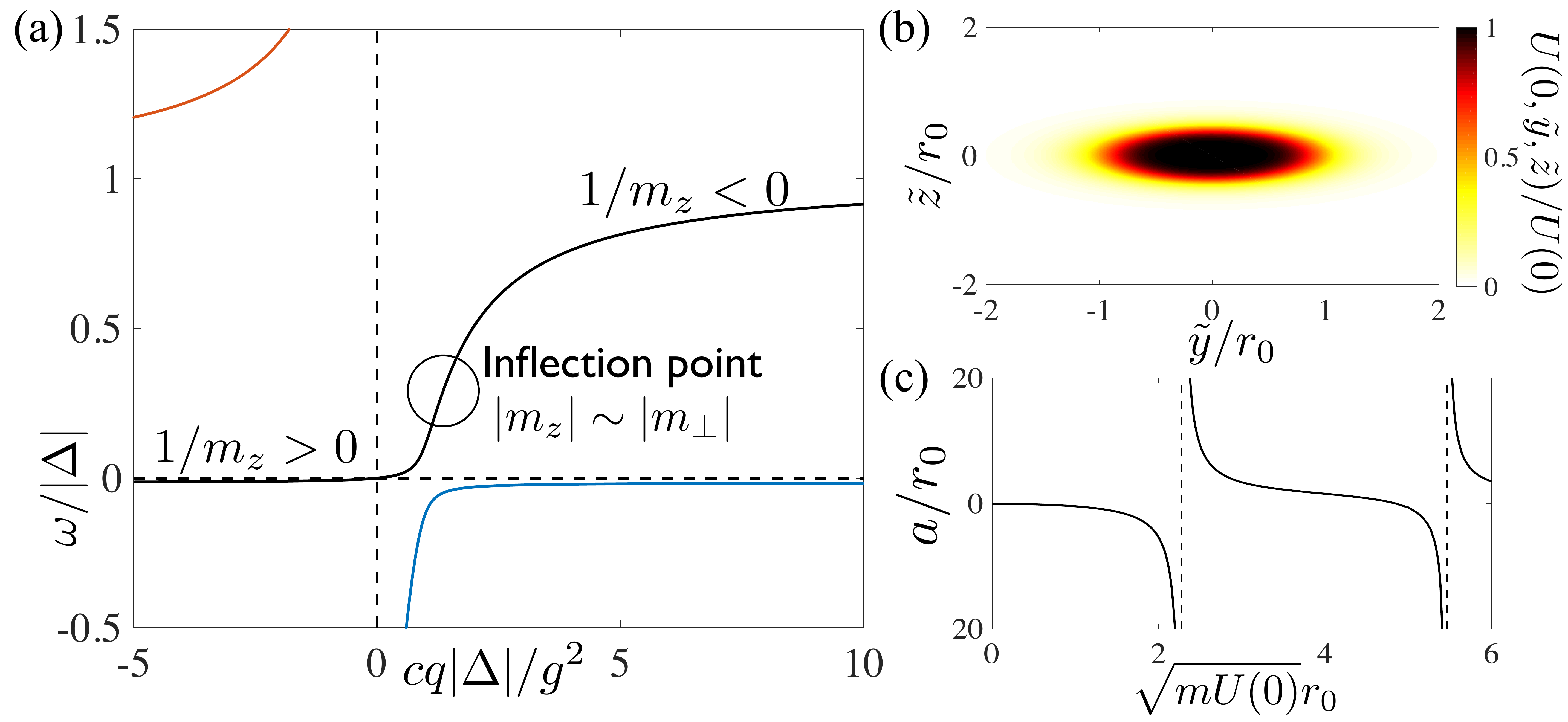}
\caption{(a)  Dispersion relation near EIT resonance and inflection point for three polariton branches  $U$ (red),  $D$ (black), and  $L$ (blue).   We took $\Omega/2\pi = 5$~MHz, $\Delta = 40$~MHz, and $g/2\pi=100$~MHz.  Here $g$ was taken to be smaller than its value in typical experiments to aid visibility.    (b) Shape of potential in rescaled coordinates. (c)  Scattering length as a function of the depth of the potential for $m_\perp/m_z =10$. The first  resonance occurs near $\sqrt{m U(0)} r_0 = 2.27$.  }
\label{fig:anismass}
\end{center}
\end{figure}

\emph{Few-Body Scattering with Anisotropic Mass.---}To understand the origin of the non-interacting part of Eq.~(\ref{eqn:H}), we consider the Hamiltonian for a single polariton with total wavevector $k= \sqrt{(k_0+q_z)^2+q_\perp^2} $  $(\hbar=1)$
\be
H=\left( \begin{array}{c c c }
ck-c k_0 & g &  0\\
g& \Delta & \Omega \\
0 & \Omega & 0
\end{array}\right),
\ee
where $\bq$ is the momentum relative to $k_0 \hat{z}$, $g=\mu_{\rm at}\sqrt{ c k_0 n/\hbar \epsilon_0}$ is the single-photon Rabi frequency of the probe, $\epsilon_0$ is the dielectric constant, $n$ is the atomic density, $\mu_{\rm at}$ is the atomic dipole moment, $\Omega$ is the control field Rabi frequency, and $\Delta$ is the detuning between the control field and $\ket{p}$ to $\ket{s}$ transition frequency [see Fig.~\ref{fig:cartoon}(a)].  We  include the decay from the intermediate state by adding an imaginary component to $\Delta \to \Delta - i \gamma$, where $\gamma$ is the halfwidth of the $p$-state.    For every $\bq$, $H$ has three eigenvalues $\epsilon_\mu(\bq)$, which can be used to find the group velocity and effective mass of the polaritons
\begin{align}
v_g&= \frac{d \epsilon_{\mu^*}}{d q_z}\Big|_{\bq^*} = c\frac{[\Omega^2 + \omega (\Delta-\omega) ]^2}{g^2(\Omega^2+\omega^2)} ,\\ \label{eqn:mz}
\frac{1}{m_{z}} & = \frac{d^2 \epsilon_{\mu^*}}{d q_z^2}\Big|_{\bq^*}=    \frac{2 v_g^2}{\Omega^2+\omega^2}\frac{\Omega^2 ( \Delta-3 \omega)-\omega^3}{\Omega^2 + \omega(\Delta-\omega)} , \\
\frac{1}{m_\perp}& =\frac{d^2 \epsilon_{\mu^*}}{d q_\perp^2}\Big|_{\bq^*}= \frac{v_g}{k_0}, 
\end{align}
where $\epsilon_{\mu^*}(\bq^*)=\omega$ is only satisfied for one choice of $q^*$ and $\mu^* \in \{U,D,L\}$ [see Fig.~\ref{fig:anismass}(a)] and we take $\bq^*_\perp=0$.   
Here we have neglected higher order corrections in $\Omega/g$.  From these expressions we see that, on EIT resonance ($\omega=0$), the mass ratio $m_z/m_\perp = g^2/2 c \Delta k_0=  (3 \pi \gamma/ \Delta) n k_0^{-3}$.

More generally, the mass ratio can be independently tuned by taking advantage of the unconventional dispersion relation for the dark-state polaritons.  This is illustrated in Fig.~\ref{fig:anismass}(a), which shows that the inverse longitudinal mass goes through a sign change for incoming probe frequencies away from the EIT resonance.  From Eq.~(\ref{eqn:mz}) we see that the inflection point occurs near $\omega=(\Omega^2 \Delta)^{1/3}$.   Operating near this inflection point allows one to equalize the mass ratio; however, it does not remove the effect of inelastic losses.  

We define the average mass $m^{-1} = (m_z^{-1}+m^{-1}_\perp)/2$ and parametrize the mass ratio as $ \tan^2(\beta) = m_z/m_\perp$.  
The condition to neglect the inelastic losses, which applies to both the EIT resonance and the inflection point in the limit $\Delta \gg \gamma,|\omega|$, is given by 
\be \label{eqn:highdens}
\frac{\textrm{Re}(m^{-1})}{\textrm{Im}(m^{-1})} \approx \frac{3 \pi }{2 \beta^2} \frac{n}{k_0^{3}} \gg 1.
\ee
For example, for atomic densities near $10^{13}~$cm$^{-3}$, $\beta$ should be less than $0.1$. It is also worth noting that the regime where $\beta \approx 1$ and Eq.~(\ref{eqn:highdens}) predicts small inelastic losses precisely coincides with the regime of Dicke super-radiance $n/k_0^3 \gtrsim 1$ \cite{Dicke54}.  In this regime, our assumption of independent decay channels for the atomic radiation would have to be re-visited.

To understand the role of such a large anisotropy in the mass on the few-body scattering problem, we first consider the two-body problem with an anisotropic mass.  
The Schr{\" o}dinger equation for two particles in the center of mass frame takes the form
\be \label{eqn:anismass}
-\frac{1}{m}\tilde{\nabla}^2  \psi + U(\tilde{\bm{r}}) \psi = E \psi
\ee
where $E$ is the energy and we have transformed to  rescaled coordinates $\tilde{z} = z / \sqrt{2} \cos \beta$ and $(\tilde{x},\tilde{y}) = (x,y)/\sqrt{2} \sin \beta$ such that the kinetic energy term becomes isotropic and the interactions become anisotropic [see Fig.~\ref{fig:anismass}(b)].  The characteristic length scale of the potential in the rescaled coordinates is given by $r_0 = r_b/\sqrt{2}\sin \beta$.
 In these rescaled coordinates we see that the interaction term mixes different two-body angular momentum $\ell$ sectors.  

For low-energies, however, the higher angular momentum channels are subject to a large centrifugal barrier, which allows the interaction terms that mix angular momentum sectors to be treated perturbatively.  
In particular, for an interaction potential that falls off as $1/r^\delta$ with $\delta>3$ and for $\ell+\ell' \geq 2$, the scattering-matrix elements (so-called $T$-matrix) scale as \cite{Sadeghpour00,Bohn09}
\be \label{eqn:T}
|T_{\ell \ell'}^{(m){\rm}}| \sim {\rm constant}\, k^{\ell + \ell' +1} + {\rm constant}\, k^{\delta-2}.
\ee
We numerically verify these scalings near threshold for a large value of the mass ratio in the supplemental material \cite{supp}.
These scalings suggest that the potential appears completely isotropic near threshold as all the partial waves beyond the $s$-wave channel $\ell,\ell' >0$ are suppressed in the absence of higher-partial wave resonances.  
%On the other hand, for the $s$-wave channel there is no centrifugal barrier and, even near threshold, the non-perturbative corrections to Eq.~(\ref{eqn:T}) are important.  
%\be
%\sigma_{\rm Born} = \frac{\pi}{k^2} \sum_{\ell \ell' m} |T_{\ell \ell'}^{(m),{\rm Born}}|^2 \approx  \pi |\Gamma_{00}^{(0)}|^2 + O(k^4).
%\ee
To tune near unitarity in this system we take advantage of shape resonances in the $s$-wave scattering length $a$.
In Fig.~\ref{fig:anismass}(c) we show the positions of the first two scattering resonances as a function of the depth of the potential $\sqrt{m U(0)}r_0$, which can be tuned via  $\Omega$ or $\Delta$.

\begin{figure}[t]
\begin{center}
\includegraphics[width=.49 \textwidth]{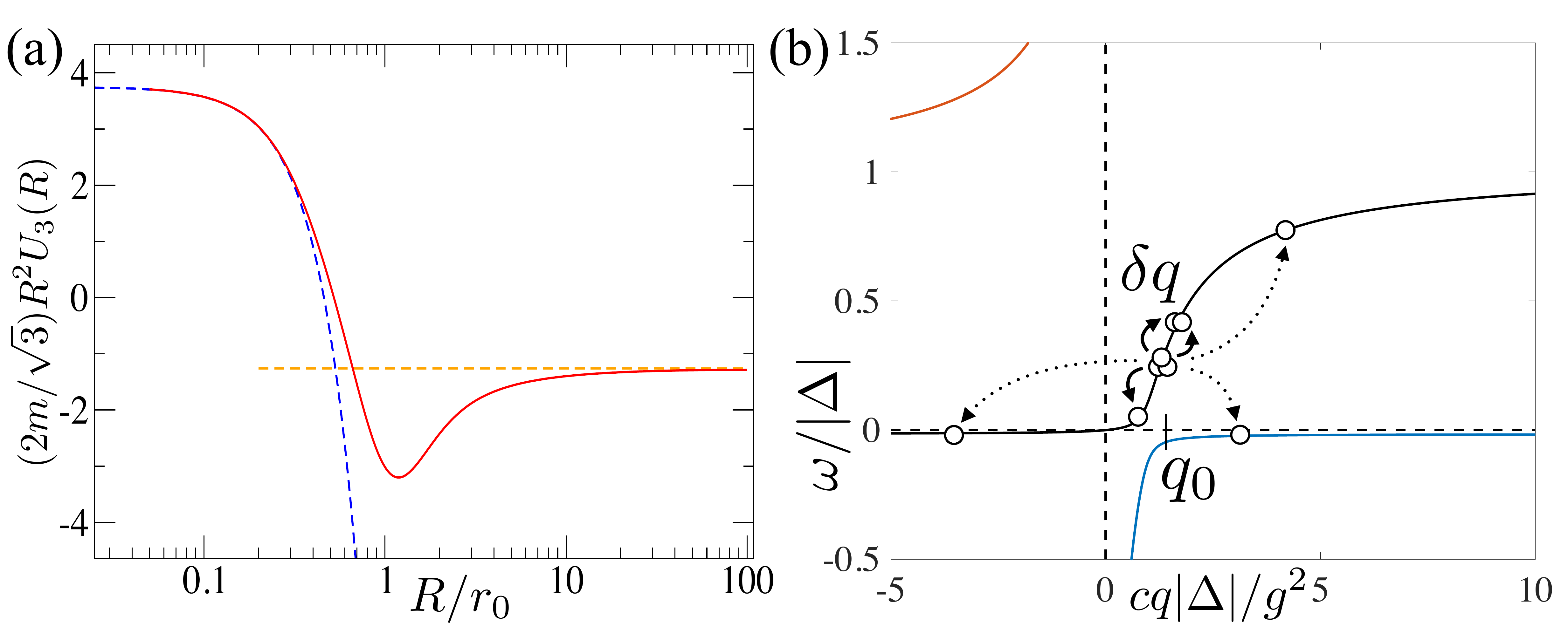}
\caption{(a) (red) Rescaled three-body hyperspherical potential at the first scattering resonance for $m_\perp/m_z=10$.  (yellow) In the region $R \gg r_0$, $ U_3(R)$ approaches the universal form that gives rise to the Efimov effect.  (blue) Analytic result in the region $R \ll r_b$. (b)  Three-body loss processes near inflection point (same branch and different branch, see text), where $q_0$ is the incoming momentum of the polaritons.    
%(b) Spectrum of Efimov states in terms of $K=-\sqrt{m|E|}$.  When the inverse scattering length is tuned to the values where the trimers reach zero energy, the Efimov states will appear in the transmission through the medium.
}
\label{fig:recomb3d}
\end{center}
\end{figure}

These features of the two-body problem have important implications for the three-body problem as well.  In particular, this analysis directly implies that  the three-body hyperspherical potential $U_{3}(R)$ will have the universal behavior in the region $ r_0 \ll R \ll a$ \cite{Efimov71,Braaten06}
\be \label{eqn:efimov}
- \frac{\sqrt{3}}{2 m} \bigg[ \frac{d^2}{d R^2} + \frac{ s_0^2+1/4}{  R^2} \bigg] f_n(R) = E_n  f_n(R)
\ee
where $s_0= 1.00624\ldots$, $R$ is the three-body hyperradius in the rescaled coordinates \cite{supp}, and $f_n$ is the hyperradial component of the three-body wavefunction in hyperspherical coordinates.  For distances on the order of $r_0$ this equation is no longer accurate as the character of the two-body interactions [Eq.~(\ref{eqn:u})] become important.  
However in the region, $R\ll r_b$ we can also find the hyperspherical potential analytically because the two-body potential  approaches a constant.
% and
%\be
%-\frac{\sqrt{3}}{2 m} \bigg[  \frac{d^2}{dR^2} - \frac{15/4}{R^2} \bigg] f_n(R) = [E_n-3 U(0)] f_n(R),
%\ee
%which has a repulsive centrifugal barrier.    
To better understand the intermediate region $r_b \lesssim R \lesssim r_0$, we have performed numerical calculations \cite{Dincao05,Dincao11} of $U_3(R)$ that include the coupling to higher-partial waves as a perturbative correction to the two-body $s$-wave potential \cite{supp}.  The results are shown in Fig.~\ref{fig:recomb3d}(a) for the first scattering resonance with $m_\perp/m_z=10$.  We see good agreement with the two analytic predictions in the small and large $R$ limits.  
The presence of the long-range $1/R^2$ potential near unitarity shows that the three-body problem will give rise to an Efimov effect with an infinite series of three-body bound states, with energies $(n=0,1,\ldots)$
\be
E_n = -\frac{\kappa_*^2}{m} e^{-2\pi n/s_0},
\ee
while the presence of the centrifugal barrier near $R \sim r_0$ suggests the scaling for the three-body parameter $\kappa_* \sim 1/r_0$ \cite{Wang11,Wang12,Naidon14}.

\emph{Suppression of Three-Body Loss.---}In cold-atom systems, the lifetime of the Efimov states is limited by their decay into deeply bound two-body states.  In this Rydberg polariton system we can avoid such effects by tuning the system near the first scattering resonance in Fig.~\ref{fig:anismass}(c), where no deep two-body bound states exist.  We can also avoid inelastic two-body loss by going to sufficiently large detunings $\Delta$.  However, due to the multi-component nature of the polariton system and the unconventional dispersion relations, three-body loss processes are still allowed whereby energy and momentum is conserved by scattering the polaritons far away from their initial momentum.  As shown in Fig.~\ref{fig:recomb3d}(b), near the inflection point, two such three-body loss processes are allowed.  In one case, one of the polaritons has a final state on the lower polariton branch, while, in the other case, all three polaritons end on the dark-state branch.  On EIT resonance, energy and momentum conservation only allow the former process.

Despite the presence of these additional loss channels, we find that their contribution to the three-body loss is  strongly suppressed because the Rydberg blockade mechanism leads to an exponentially small two-body potential for large relative momenta $\delta q \gg r_b^{-1}$. 
 In particular, for the  three-body loss channels  in Fig.~\ref{fig:recomb3d}(b) each process  involves a finite momentum transfer $\delta q$.   When the minimum $\delta q \gg r_b^{-1}$, we can estimate the three-body loss rate perturbatively in the Fourier transform of $U(r)$, which, it is readily seen, is exponentially suppressed as $ e^{- \delta q r_b}.$
%\be
%U_{\bq}=- \frac{4 \pi^2 \alpha\, r_b^2}{3 q \bar{\chi}} \bigg[ e^{-q r_b} - 2 \textrm{Re} \Big(e^{- q r_b e^{i \pi/3} -i\pi/3}\Big) \bigg].
%\ee
%From which we see that the three-body loss channels will be exponentially suppressed  as $e^{- \delta q r_b}$.  
Near the inflection point, the minimum $\delta q$ (and thus the dominant channel) occurs for the process where all three polaritons end on the dark-state branch.  
By analytically by expanding the dispersion to third order around the momentum of the polaritons, we find that the
%$\bq_0$
%\be
%\epsilon_D(q) = \epsilon_D(q_0)+v_g \delta q_z + \frac{\delta q_z^2}{2 m_z}+\frac{\delta q_\perp^2}{2 m_\perp} + \frac{\eta \delta q_z^3}{6},
%\ee
%where $\eta=d^3 \epsilon_D/d^3q_z|_{\bq_0}$.
%The three-body resonance condition $3\epsilon_D(\bq_0)=2\epsilon_D(\bq_0+\delta \bq) +\epsilon_D(\bq_0-2\delta \bq)$ has a minimum $|\delta \bq|$ when $\delta q_\perp=0$ and $\delta q_z=3/\eta m_z$.  
three-body losses will be suppressed when
\be \label{eqn:lossreq}
\delta q_z r_b \approx \bigg(\frac{g^2}{c k_0 \Delta } \bigg)^{5/3}  ( r_b k_0)^{1/3} \frac{\Phi^{2/3}}{\beta^3} \gg1,
\ee 
where the value of $\Phi \equiv \sqrt{ m U(0)}r_0$ is determined by the position of the first scattering resonance [e.g., see Fig.~\ref{fig:anismass}(c)].    In the supplemental material we provide a more detailed discussion of the scaling of the three-body loss parameter \cite{supp}.  For a density of $10^{13}~$cm$^{-3}$ with a blockade radius of $20~\mu$m \cite{Firstenberg13}, this implies that $\beta$ should be less than $0.1$ to strongly suppress the three-body losses.  

On EIT resonance, the minimum $\delta q \approx g^2/c \Delta$ and the condition to neglect three-body loss is simply that $\phi \gg 1$.  This scaling implies  that three-body loss will still play an important role near the first scattering resonance because, on EIT resonance, $\Phi \approx \phi$.  Thus we see the primary advantage of working near the inflection point is the control it provides over the mass ratio, scattering length, and three-body loss rate.  

%In the supplement, we present a four-level scheme that provides  independent tuning of the mass ratio at the EIT resonance \cite{supp}

\emph{Preparation and Detection.---}To prepare the Efimov states we propose to use the high-degree of control over the two-body parameters to tune the scattering length ($a<0$) to values in which the trimers cross zero energy.  As each such value is crossed in the space of control paramters, an Efimov state will emerge in transmission through the medium due to the quench dynamics described in the introduction.  When more than one Efimov state is present in the medium, they can be distinguished from each other by changing the transverse focus of the input field to increase the initial overlap with the desired  state.  

The two primary requirements to realize such long-lived Efimov trimers are given by Eq.~(\ref{eqn:highdens}) and Eq.~(\ref{eqn:lossreq}).  Three other important considerations are  (i) the size of the Efimov state does not exceed the size of the atomic cloud, (ii) the atomic density is low enough to avoid molecular Rydberg state effects \cite{Balewski13,Komar15,Thompson17,Busche17} and (iii) the thermal motion of the atoms does not break up the Efimov  states.  The first requirement is of particular relevance for small values of $\beta$ because, when the coordinates are transformed back to the lab-coordinates, we see that the longitudinal size of the Efimov states will scale as $r_0 \approx r_b/\beta$.  For an atomic density of $10^{13}~$cm$^{-3}$, we require $\beta < 0.1$, which implies that the length of the atomic cloud should be greater than 100~$\mu$m.  
%The condition to observe the second Efimov trimer is more demanding as this state is a factor of 22.7 times larger.  However, this scale factor can be reduced by using mixed species systems with differing mass ratios \cite{Dincao06}, which is achievable in our system by using multi-color input fields.  
For these densities the mean inter-atomic spacing is around $0.5$~$\mu$m.  To avoid collisions between the Rydberg electron and other ground state atoms, this places an upper bound on the Rydberg quantum number in the range of $n=50-100$.  In the case of ${}^{87}$Rb this leads to $r_b \approx 5-15$~$\mu$m, for which the first polariton scattering resonance [see Fig.~2(c)] is readily achievable at these densities.      To neglect motional dephasing of the atoms we require that the doppler broadening of the two-photon transition is much less than the binding energy of Efimov states.  For the case of ${}^{87}$Rb at a temperature of 35~$\mu$K, the doppler broadening is on the order of 100~kHz, while the Efimov binding energy scales as $\Omega^2/\Delta$, which can be much larger than 1~MHz depending on parameters.  More generally we remark that the low-energy nature of the Efimov states will render them insensitive to many corrections arising from short-range physics; thus we expect their emergence to be a robust feature of the Rydberg polariton system whenever it is possible to tune the system near a resonance in the 3D two-body scattering length.

\emph{Conclusions.---}We have demonstrated that systems of interacting photons formed from Rydberg polaritons naturally give rise to an Efimov effect.  A potential advantage of this system is that one can realize long-lived Efimov trimers by tuning the system to the first scattering resonance and suppressing other three-body loss channels.  The wide range of control over the system parameters and the ability to supress $N$-body losses make this a promising platform for studying few-body universality. At the same time,  increasing the input light intensity provides access to the non-equilibrium, many-body regime where the role of universal few-body physics is poorly understood.   The resulting photonic states that emerge from the medium have a rich multi-particle entanglement structure, which may enable them to be used as a resource for  optics-based quantum technologies \cite{Pan12,Shalm13}.

\begin{acknowledgements}
\emph{Acknowledgements.---}
We acknowledge helpful discussions with  W. D. Phillips, T. Porto, H. P. B{\" u}chler, P. Zoller, M. D. Lukin, V. Vuletic, and Q.-Y. Liang.  MJG and SD would like to thank the organizers of the workshop on ``Quantum Simulation and Many-Body Physics  with Light'' in Crete, Greece where some of this work was completed.  This research was supported by the  European Research Council through Grant Agreement No. 647434 (DOQS), NSF through Grant No. PHY-1607204, ARL CDQI, NSF QIS, AFOSR, ARO, ARO MURI, and  the NSF-funded Physics Frontier Center at the JQI.
\end{acknowledgements}

\bibliographystyle{../../apsrev-nourl-title-PRX}
\bibliography{../RydUniv.bib}

\pagestyle{empty}
%\foreach \x in {1,2,3,4,5}
{ 
\begin{figure*}
\vspace{-1.8cm}
\hspace*{-2cm} 
\includegraphics[page=1]{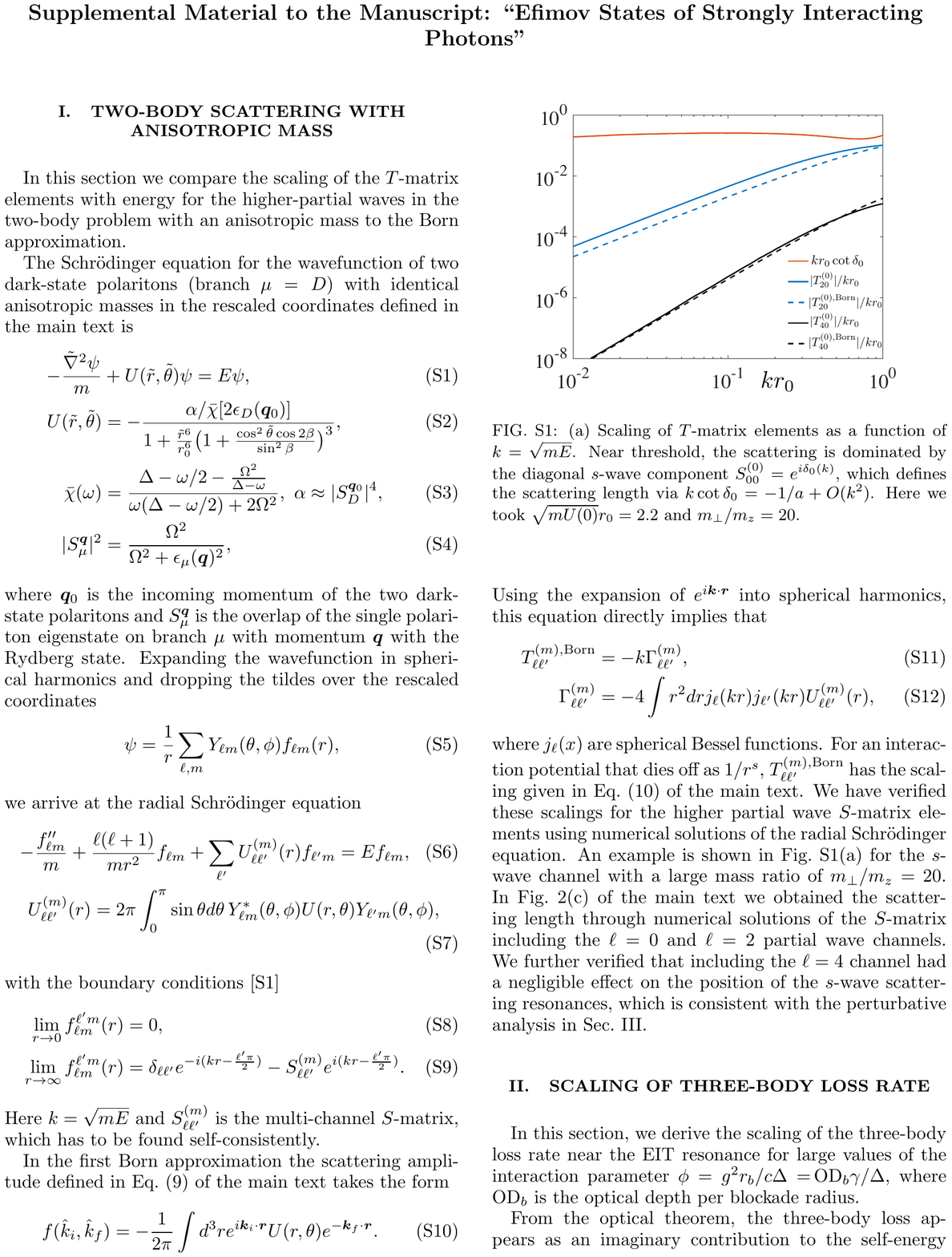}
\end{figure*}

\begin{figure*}
\vspace{-1.8cm}
\hspace*{-2cm} 
\includegraphics[page=2]{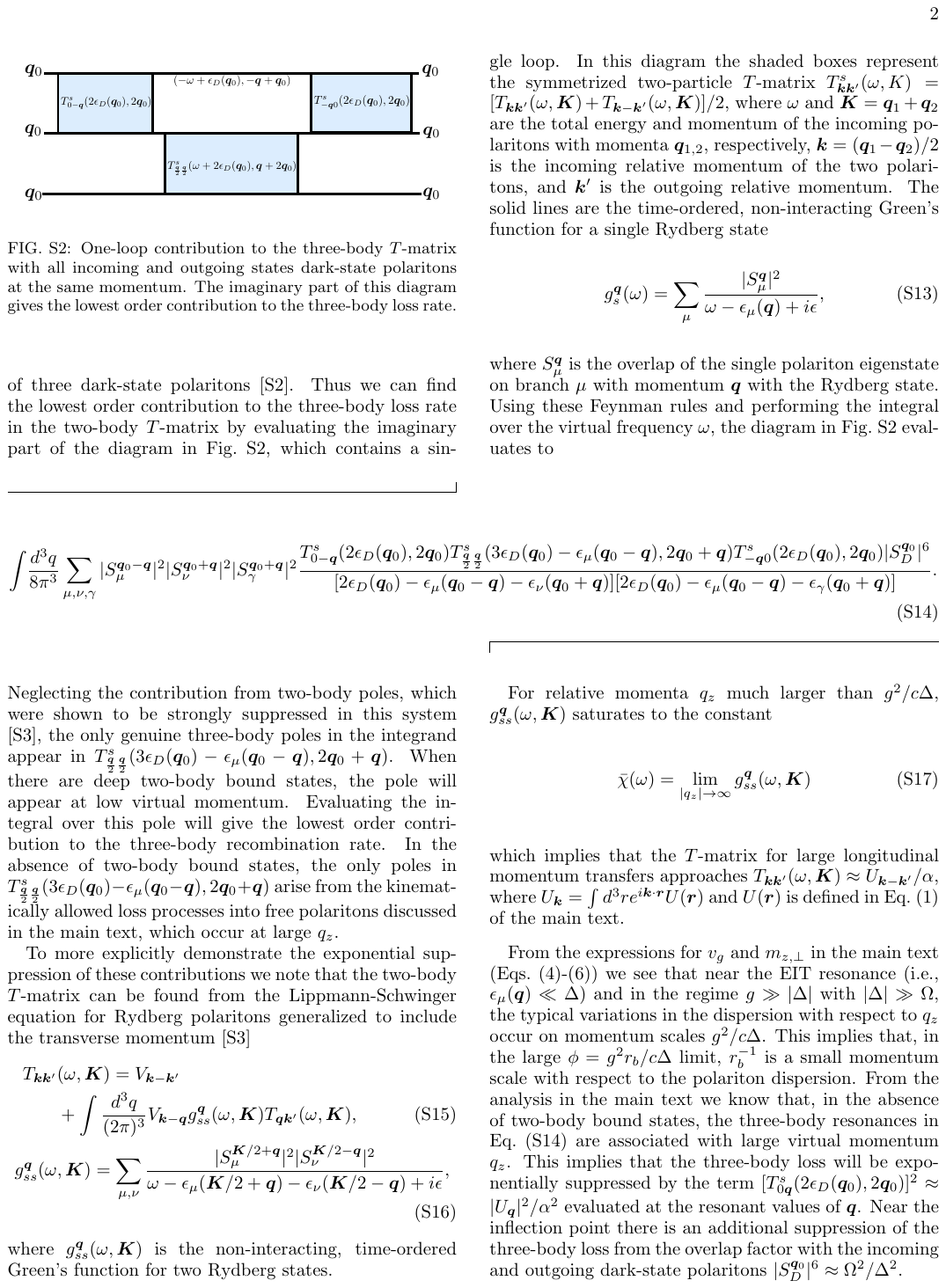}
\end{figure*}

\begin{figure*}
\vspace{-1.8cm}
\hspace*{-2cm} 
\includegraphics[page=3]{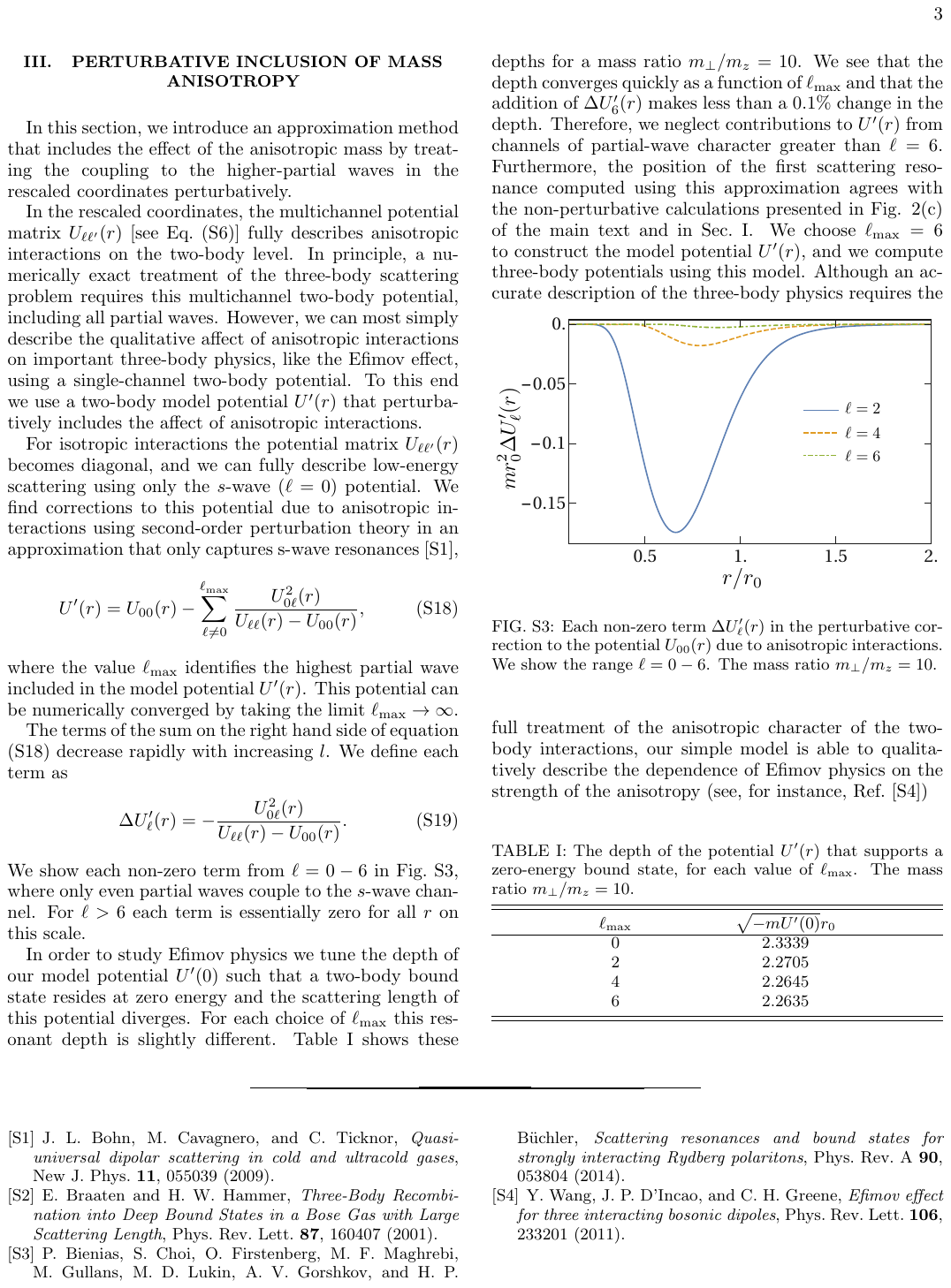}
\end{figure*}

%\begin{figure*}
%\vspace{-1.8cm}
%\hspace*{-2cm} 
%\includegraphics[page=4]{RydEfimov_supp.pdf}
%\end{figure*}
%
%\begin{figure*}
%\vspace{-1.8cm}
%\hspace*{-2cm} 
%\includegraphics[page=5]{RydEfimov_supp.pdf}
%\end{figure*}
}

\end{document}